\documentclass[%
 reprint,
 amsmath,amssymb,
 aps,
 pra,
]{revtex4-1}

\usepackage{mwe}
\usepackage{graphicx}
\usepackage{dcolumn}
\usepackage{appendix}
\usepackage{bm}
\usepackage{todonotes}
\usepackage[caption=false]{subfig}
\DeclareMathOperator{\Tr}{Tr}

\newcommand{\figref}[1]{\ref{#1}}

\usepackage{algorithmic}
\usepackage{algorithm}
\renewcommand{\algorithmiccomment}[1]{\hfill$\triangleright$ #1}

\usepackage[normalem]{ulem}
\usepackage{color}

\begin{document}

\preprint{APS/123-QED}

\title{Adaptive Quantum State Tomography with Neural Networks} 
\author{Yihui Quek (equal contributions)}
  \email{yquek@stanford.edu}
\affiliation{Department of Applied Physics and Information Systems Laboratory, Stanford University}
 \author{Stanislav Fort (equal contributions)}
  \email{sfort1@stanford.edu}
 \affiliation{Department of Physics, Stanford University}
\author{Hui Khoon Ng}
\email{cqtnhk@nus.edu.sg}
\affiliation{Yale-NUS College, Singapore}
\affiliation{Centre for Quantum Technologies, National University of Singapore}
\affiliation{MajuLab, CNRS-UCA-SU-NUS-NTU International Joint Research Unit, Singapore}

\date{\today}

\begin{abstract}
Quantum State Tomography is the task of determining an unknown quantum state by making measurements on identical copies of the state. Current algorithms are costly both on the experimental front -- requiring vast numbers of measurements -- as well as in terms of the computational time to analyze those measurements. In this paper, we address the problem of analysis speed and flexibility, introducing \textit{Neural Adaptive Quantum State Tomography} (NA-QST), a machine learning based algorithm for quantum state tomography that adapts measurements and provides orders of magnitude faster processing while retaining state-of-the-art reconstruction accuracy. Our algorithm is inspired by particle swarm optimization and Bayesian particle-filter based adaptive methods, which we extend and enhance using neural networks. The resampling step, in which a bank of candidate solutions -- particles -- is refined, is in our case learned directly from data, removing the computational bottleneck of standard methods. We successfully replace the Bayesian calculation that requires computational time of $O(\mathrm{poly}(n))$ with a learned heuristic whose time complexity empirically scales as $O(\log(n))$ with the number of copies measured $n$, while retaining the same\ reconstruction accuracy. This corresponds to a factor of a million speedup for $10^7$ copies measured. We demonstrate that our algorithm learns to work with basis, symmetric informationally complete (SIC), as well as other types of POVMs. We discuss the value of measurement adaptivity for each POVM type, demonstrating that its effect is significant only for basis POVMs. Our algorithm can be retrained within hours on a single laptop for a two-qubit situation, which suggest a feasible time-cost when extended to larger systems. It can also adapt rapidly to a subset of possible states, a choice of the type of measurement, and other experimental details.\\

\end{abstract}


\maketitle

\section{\label{sec:Intro}Introduction}

Quantum state tomography (QST) is the task of estimating the density matrix of an unknown quantum state, through repeated measurements of the source, assumed to put out identical copies of the state. This procedure can be used to characterize not only quantum states, but also processes acting on quantum states, and is an indispensable subroutine in quantum information processing tasks (see, for example, Ref.~\cite{QST_text}, for a review). QST is, however, a resource-heavy task, as one needs many measurements on many copies of the quantum state, to yield sufficient data for a good estimate of the $d^2-1$ real parameters needed to describe the state of a $d$-dimensional quantum system. For estimating quantum processes, the number of parameters needed is $d^4$, a much worse scaling. Much of the research in QST is hence about minimizing the resource cost of getting an estimate with a target precision.

One angle of attack is to use adaptive measurements, adjusting the measurement to be done on the next copy of the state, based on the information gathered from the copies measured so far, to maximize (according to some chosen measure) the information gained from that next copy. One approach was proposed in \cite{Fischer-Quantum-statemeasurements} (under the name \textit{self-learning}) and generalized in \cite{HuszarAdaptiveTomography} as Adaptive Bayesian Quantum Tomography (ABQT). These methods were later experimentally implemented in \cite{Hannemann2002Self-learningStates,Struchalin2016ExperimentalStates}. In this paper, we propose a neural network-aided alternative to adaptive Bayesian particle-filter based schemes in the vein of ABQT \cite{HuszarAdaptiveTomography}. We call our scheme \textit{Neural Adaptive Quantum State Tomography} (NA-QST). 
 
 At its core, ABQT uses Bayes' rule to update the prior distribution on the quantum state space to the posterior according to the outcome of measurements done so far. Then, the adaptation is done using a merit function computed as an integration over the current posterior distribution. In order to make this integration numerically tractable, the state space is represented by discrete samples, known as \textit{particles}. Accompanying each sample is a \textit{weight} that represents its relative likelihood, which is sequentially updated in a Bayesian fashion based on measurement results. The \textit{adaptivity} comes from the option to choose subsequent measurement configurations to maximize the expected information gain, given the updated posterior distribution. 

A practical issue encountered in filtering algorithms is the gradual decay of the vast majority of particle weights, which undercuts the efficacy of the weight updates. In ABQT, and other Bayesian-type tomography procedures, this is a particularly acute problem since the likelihood function that enters the posterior distribution typically becomes very sharply peaked as the number of copies measured grows. The numerical fix is to resample the bank periodically (i.e. the particles and weights must be chosen anew). As the last step of resampling, \cite{Struchalin2016ExperimentalStates} proposes to use the Metropolis-Hastings algorithm to \textit{mutate} the particles, or perturb them isotropically in state space. However, running the Metropolis-Hastings step is very computationally intensive, due to the need to perform a number of computations linear in the number of previous measurement results to determine the acceptance probability. Since the resampling must happen before the next measurements are determined, this significantly prolongs the overall runtime of the adaptive QST algorithm.

In this paper, we demonstrate a custom-built recurrent neural network architecture that learns to perform quantum state tomography directly from data. Our approach automatically develops an efficient resampling and mutation step to refine the candidate solutions, as well as a way of incorporating new measurements to its current best estimate of the state. Our algorithm also learns to predict the suitable next measurement that, if performed, would lead to the most information gained given its current knowledge about the state. We implemented the algorithm in an end-to-end differentiable manner, allowing us to train all of its components jointly.  

The efficiency of our approach comes from using machine learning to learn an approximate replacement of the Bayesian update rule to learn new weights on the particles after every batch of measurements. This eliminates the problem of weight decay, and therefore, the need for time-costly resampling. Although our weights (unlike in ABQT) no longer have the interpretation of being the posterior distribution, they are just as effective as inputs to the adaptive step: to compute a heuristic which is maximized to choose the next measurement. All in all, our adaptive algorithm matches the performance of ABQT, while speeding it up significantly, giving a practical, genuinely on-the-fly adaptive QST scheme. Our NA-QST algorithm enjoys the same scaling of accuracy with number of measurements as ABQT, a state-of-the-art adaptive algorithm, but with a computational speedup of up to a million (for $10^7$ measurements) when run on the same hardware. Our approach is furthermore agnostic to the number of qubits involved, and the type of measurements used, and can be retrained within hours on a single laptop to suit particular experiment's details. During the training phase, our algorithm runs a simulation of quantum mechanical measurements to substitute physical measurement results, that are to be provided during the algorithm's deployment after training.  

Some aspects of our approach can be generalized beyond the field of quantum tomography. We have developed a fully differentiable (in the machine learning sense of the word, i.e.,exactly differentiable using existing machine learning libraries) implementation of measurement in quantum mechanics, which more generally can simulate any experiment with probabilistic outcomes in TensorFlow. In addition, a technical takeaway from this work is that when there are disjoint components in an algorithmic pipeline -- in this case, particle filters that pass weights to the objective function optimized to choose the next measurement -- an end-to-end machine learning pipeline can be trained jointly to integrate the steps and have one part of the system producing optimal results for the next. This is of interest to the many areas of Physics and Engineering to which particle filters have been applied \cite{vanLeeuwen2009ParticleSystems,Vermaak2005SequentialMC}.

\section{Neural Network-Based Adaptive Quantum State Tomography (NA-QST)}

\subsection{Basic definitions for QST}\label{sec:QSTDef}
We define here some basic terminology needed for discussing the problem of QST. As mentioned earlier, the task of QST is to estimate the state, i.e., the density matrix, $\rho$, of a quantum system. $\rho$ is then a trace-1 and Hermitian operator on the $d$-dimensional Hilbert space $\mathcal{H}$ of the quantum system, represented by a $d\times d$ matrix. We assume that we have access to a source that puts out independent and identical copies of the (unknown) state $\rho$. We are allowed to make measurements on the state, and from the gathered data, estimate $\rho$. Generalized measurements, also often referred to as positive operator-valued measure (POVM), are permitted. These are describable as a set of outcome operators $\Pi\equiv \{\Pi_y\}$ on $\mathcal{H}$, satisfying $\Pi_y \geq 0 \, \, \forall y$ and $\sum_y\Pi_y=\mathbf{I}$, the $d$-dimensional identity operator.

For a chosen $\Pi$, the probability that one gets a click in the detector corresponding to the outcome $\Pi_y$ is given by the Born's rule,
\begin{equation}
p_y=\mathrm{Tr}(\rho\Pi_y) \, .
\end{equation}
The likelihood for getting data $\mathcal{D}$, a sequence of detector clicks, summarized by $\{n_y\}$, where $n_y$ is the total number of clicks in the detector for $\Pi_y$, is given by
\begin{equation}\label{eq:BornRule}
p(\mathcal{D}|\rho)=\prod_y p_y^{n_y} \, ,
\end{equation}
for $p_y$s computed from $\rho$ by the Born's rule.
$p(\mathcal{D}|\rho)$ is a probability distribution over the data, i.e., $\sum_\mathcal{D}p(\mathcal{D}|\rho)=1$.

We let $\nu_y\equiv n_y/N$, where $N\equiv \sum_yn_y$ is the total number of copies measured. We refer to the $\nu_y$s as the relative frequencies for the different outcomes $\Pi_y$. The relative frequencies $\nu_y$ are estimates of the probabilities $p_y$s, but note that while both $\nu_y$s and $p_y$s are nonnegative numbers that sum  (over the $y$ label) to 1, $p_y$s that come from a quantum state $\rho$ through the Born's rule satisfy additional constraints due to the positivity of $\rho$. As such, setting $p_y=\nu_y$ does not always yield a set of Born probabilities for a valid quantum state, and much of the task of quantum state \emph{estimation}, in converting data to a valid quantum state, is about dealing with the positivity constraints.

In Bayesian estimation procedures, one talks about the prior and the posterior distributions on the quantum state space. The prior distribution captures our initial knowledge about the identity of the state \emph
{prior} to any data-taking. We denote it as $\mathrm{d}\rho\,p(\rho)$, for some suitably chosen volume measure $\mathrm{d}\rho$ on the state space. $p(\rho)$ is the prior density, while $\mathrm{d}\rho \,p(\rho)$ is the infinitesimal probability that  the true state lies in the volume $\mathrm{d}\rho$, according to our prior expectations. $p(\rho)$ satisfies $\int\mathrm{d}\rho\, p(\rho)=1$. The posterior distribution, denoted as $\mathrm{d}\rho\,p(\rho|\mathcal{D})$ represents our updated knowledge, after obtaining data $\mathcal{D}$. The update from prior to posterior densities follows from Bayes rule,
\begin{equation}
p(\rho|\mathcal{D})=\frac{p(\mathcal{D}|\rho)\,p(\rho)}{p(\mathcal{D})} \, ,
\end{equation}
where $p(\mathcal{D})$ is the likelihood of the data $\mathcal{D}$, playing the role of the normalization constraint: $p(\mathcal{D})\equiv \int \mathrm{d}\rho\,p(\rho)p(\mathcal{D}|\rho)$.

Note that the ABQT algorithm relies on the posterior distribution to make decisions about the next measurement to make; in our NA-QST, as we explain below, this posterior distribution is replaced by the choices of weights on the particle samples made by the trained neural network.

\begin{figure*}
\subfloat[\label{fig: rnncell} Schematic of the recurrent neural network (RNN).]{%
  \includegraphics[width=.5\linewidth]{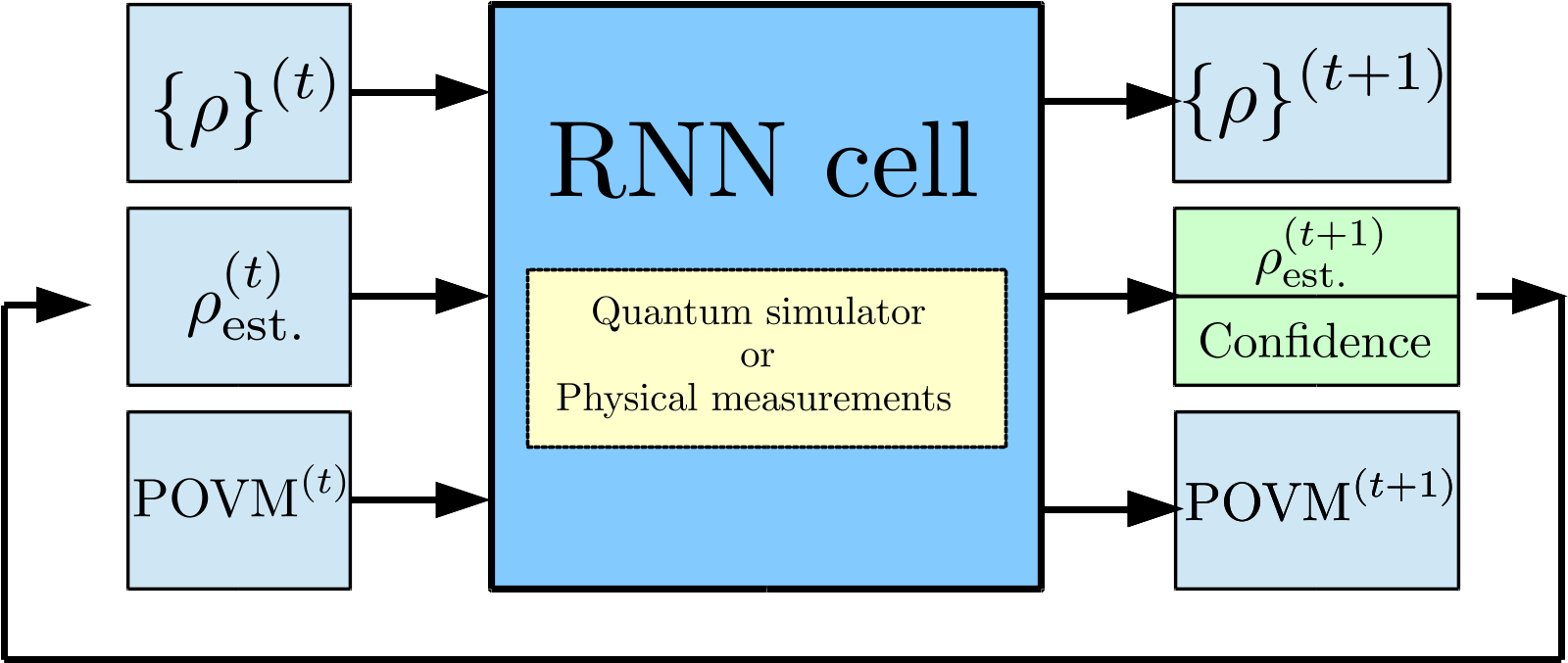}%
}\hfill
\subfloat[\label{fig: unit cell}One unit cell of the recurrent neural network (RNN), corresponding to the box labelled \textit{RNN cell} in the panel above.]{%
  \includegraphics[width=1\linewidth]{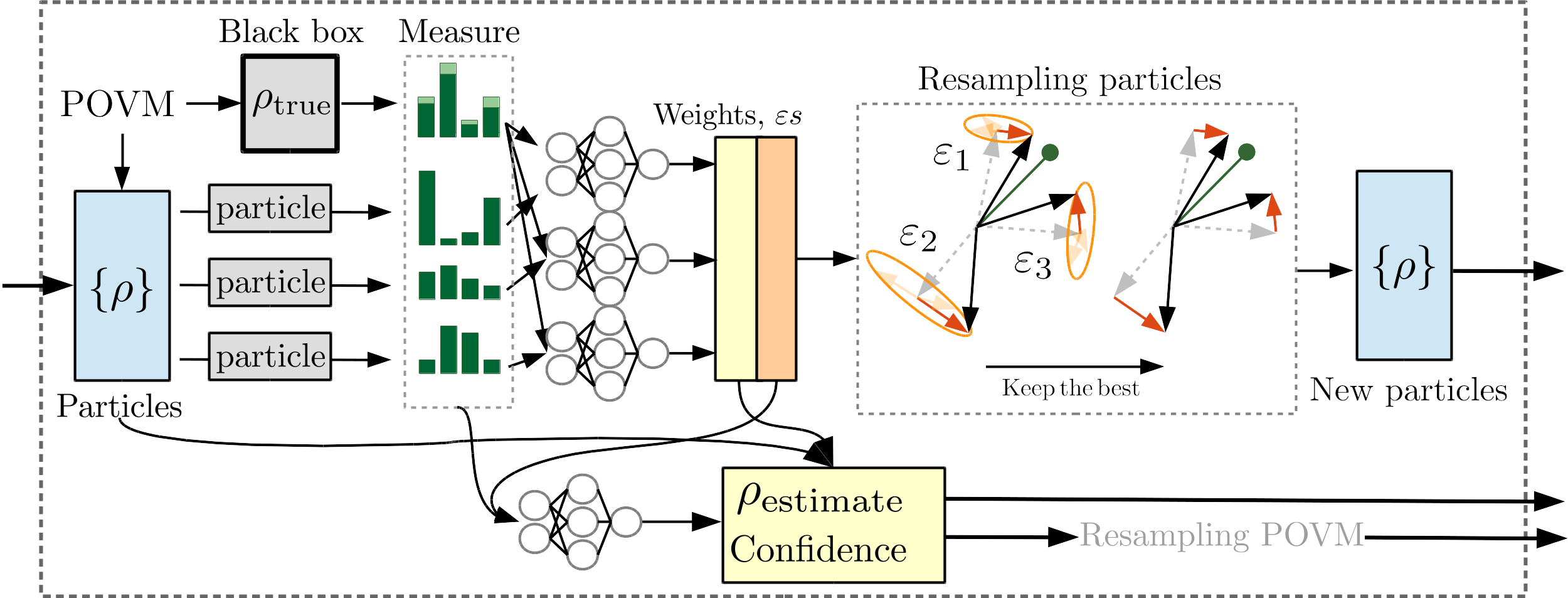}%
}
\caption{\label{fig: NN diagram}The recurrent neural network architecture. A recurrent neural network (RNN) cell takes a previously used bank of candidate solutions -- \textit{particles} -- and the best guess for the hidden density matrix as inputs, and outputs a refined bank of particles together with a new estimate of the density matrix and its internally generated confidence value. This process is successively repeated, gradually converging to the correct density matrix. During the training period, inside the RNN unit cell, a simulated quantum mechanical measurement is performed on the hidden density matrix, resulting in a noisy estimate of the distribution of the probability of measurement (POVM) outcomes. Equivalent distributions are calculated for density matrices in the particle bank. Based on $L_1$ (total variation distance) and $L_2$ (Hilbert-Schmidt distance) distances between these probability distributions, a neural network assigns weights as well as perturbation sizes $\varepsilon$ to each particle. The weights are used to construct a new estimate of the hidden density matrix, as well as its internal confidence value. Each particle is perturbed by an amount determined by the perturbation size (see main text for details). For each perturbed particle, and the probability distribution induced on it by the POVM, we calculate various distance measures on probability distributions and let a neural network decide which perturbed particle is the closest to the solution. The closest particle is kept and replaces the initial particle in the bank of particles in the next iteration. Iterating on this procedure, the candidate solutions gradually converge on the true solution. During deployment after training, the quantum measurement simulator is replaced by real physical measurements supplied by an experimenter. After each estimate, a new POVM is proposed based on maximization of a heuristic function related to entropies of the particle bank and the current estimate of the density matrix. This allows the algorithm to learn to maximize the information gained from the following measurement. The whole algorithm consists of a large number of the RNN unit cells connected in series, producing an estimate of the state at each step. The computational graph comprising a series of linked RNN unit cells is end-to-end differentiable, allowing us to train it as a single function using existing machine learning libraries. The detailed description of the algorithm is presented in Algorithm~\ref{alg:NQST-train}.}
\end{figure*}

\subsection{Machine learning and neural networks}
Machine learning has been successfully used to learn approximate solutions to a wide variety of problems. Neural networks, and deep neural networks in particular, are a class of expressive functional approximators capable of learning complicated models across many domains, ranging from image classification \cite{lecun-gradientbased-learning-applied-1998}, to game playing \cite{silver2017mastering}, to natural language understanding \cite{GoogleNMT}.

Neural networks are a class of functions that can be used as a functional \textit{ansatz} in situations where mapping of inputs to outputs is needed, while an explicit algorithm is either unknown, hard to come by, or expensive to run. By means of gradient descent, neural networks are \textit{trained} to successfully approximate an unknown function using a large number of examples and given a loss function specifying how unhappy one is with a solution. In cases where an exact input -- output mapping is available but might be expensive to run, neural networks can develop simpler, faster effective descriptions learned directly from data.

By specifying a particular neural \textit{architecture} -- a particular choice of neurons, their connections, and the way they interact -- we describe a \textit{family} of functions. The selection of a single algorithm (i.e. a single choice of these parameters) is achieved via training.

Among widely used architectures are fully-connected neural networks, convolutional neural networks, and recurrent neural networks, each of which encode priors about the function they are trying to approximate in their structure. This makes learning easier and typically leads to faster convergence to a solution. For very specific tasks (such as game playing), a very particular choice of architecture is typically made to narrow down the space of possible functions enough. In this paper, we use our knowledge of traditional quantum state tomography algorithms as well as quantum mechanics itself to construct a suitable custom-built architecture for quantum state tomography.

\subsection{Overview of our neural network algorithm}
The task of quantum state tomography is a complicated procedure, and the adaptive mapping from previous measurement outcomes to the next measurement is highly nonlinear. As such, it is hard to learn simply by feeding the raw stochastic measurement outcomes into a generic fully-connected neural network (FC-NN). 

We therefore custom-built a recurrent neural architecture that utilizes our prior knowledge of the problem, encoding it explicitly into the network structure. In order to be able to efficiently optimize our neural network, we write down the full \textit{computational graph} of the problem in \textit{TensorFlow} \cite{tensorflow2015-whitepaper} (the most commonly used library for deep learning), that in turn handles analytical differentiation internally. As a crucial part of the process, it was necessary for us to develop a differentiable implementation of quantum mechanics in TensorFlow that serves as a quantum simulator during training, mimicking real physical measurement outcomes that will be provided by an experimentalist during algorithm's deployment after training.

To train a neural network, a large number of examples and an objective/loss function are needed to determine the suitable choice of free parameters. In particular, we are making an estimate $\hat{\rho}$ of a state $\rho$ from a sequence of measurements and their outcomes, given a distance measure on density matrices. The measurements are allowed to be generalized measurements, or POVMs, as introduced earlier. The critical task is as follows: given an input (the relative frequencies of POVM outcomes), and given the POVM, output a good estimate of the state which gave rise to those measurement outcome statistics. To allow for POVM adaptivity, we perform this sequentially, enabling the subsequent estimation step to use the previous steps' estimate, as well as auxiliary useful information passed between the steps. The full approach is presented in detail in Algorithm~\ref{alg:NQST-train} and illustrated in Figure~\ref{fig: NN diagram}.

To generate training examples for the neural network, we prepared random true states to be estimated and performed simulated measurements on these states. The neural network then trains on these examples, eventually learning a sequence of functions that map the input to the output. The learning consists in tuning the weights of the neural network architecture such that the network's output is close to the true state in terms of some state distance measure. We chose the Bures distance; however, training using the simple Hilbert-Schmidt distance (Frobenius norm) works equally well. The training stops when the weights are sufficiently tuned, and no further improvement takes place.

Our algorithm allows for an iteratively improving estimate of the density matrix with every new batch of measurements. In particular, it can be stopped after any number of measurements, and since on average the distance to the true solution decreases, it provides the best estimate up to that moment. Note that even during training, our algorithm does not have direct access to the true density matrix. The only information about it comes indirectly through the relative frequencies of a finite number of simulated measurements on said state, provided by the quantum simulator we implemented. This mimics the situation during deployment in a laboratory, where the measurement data are provided by the experimenter and the true solution is unknown. 

In the testing phase, the network with tuned weights is evaluated on how well it estimates unknown states based on their empirical measurement statistics. This is exactly the same task as in the training phase, but the network can no longer adjust its weights based on the error signal coming from the knowledge of the true state. If the network performs well in the testing phase, then it is considered well functioning and can be deployed in the laboratory with its newly-tuned weights. Note that there are two notions of learning pertaining to our algorithm: the outer loop learning chooses the right algorithm, while in the inner loop the algorithm itself learns iteratively about a particular density matrix based on the measurement outcomes provided to it. In essence, we are learning to learn about density matrices.

\subsection{Detailed description of our neural network algorithm}

The architecture that we use for the task at hand is broadly known as a \textit{Recurrent Neural Network} (RNN), which features widely in Natural Language Processing tasks due to its ability to model sequences. The \textit{recurrence} comes from the fact that the network takes in outputs from previous time (iteration) steps as inputs at the current step, together with any new information available at the time. The network also maintains a \textit{memory state} that stores some information about states of the network at previous time steps. While retaining the basic common structure, we have changed the architecture of the unit cell -- the particular operation performed at every step -- completely, in order to make it suitable for working in quantum mechanical settings. We also chose specific relevant information to be passed from cell to cell, such as the bank of refined candidate solutions (particles) as well as a number of previously used POVMs and their corresponding empirical counts.

Figure \ref{fig: rnncell} shows an overall schematic of the neural network. Our algorithm shares a basic structural similarity with ABQT's particle filter: a bank, $\mathcal{B}$ of \textit{particles} $\{\rho_i\}_{i \in \mathcal{B}}$ -- that is, candidate quantum states -- and associated weights $\{w_i\}_{i \in \mathcal{B}}$. As in ABQT, our running estimate for the state is the weighted average of these particles. The key difference is the use of neural networks to determine weights over the particle bank, the particle resampling step, as well as to weigh solutions from all previous steps together into an estimate. This allows us to by-pass time consuming Bayesian calculations and replace them with simple, yet equally powerful heuristics learned directly from data. The details of the procedure are presented in Algorithm~\ref{alg:NQST-train}.

\begin{figure} 
	\begin{algorithm}[H]
		\caption{NA-QST\\Training phase with a quantum simulation}
		\label{alg:NQST-train}
		\begin{algorithmic}		
			\REQUIRE A sequence of $T$ numbers of copies to be measured $M_t$. An initial bank of $N_\mathrm{bank}$ randomly chosen valid density matrices $\left \{ \rho_i\right \}_{i \in \mathrm{bank} \mathcal{B}}$. The true density matrix $\rho_\mathrm{true}$.  
			\ENSURE A sequence of $T$ reconstructions $\{ \hat{\rho}_t \}$ and their associated losses. Overall reconstruction loss to use in training.
			\FOR{time step $t$ from 0 to $T-1$}
			\STATE Generate a random POVM given a specification if operating in random POVM mode, or use the adaptively chosen POVM coming from the previous step.
			\STATE Simulate $M_t$ measurement outcomes on the true density matrix drawn from $p(\rho_\mathrm{true},\mathrm{POVM})$. This gives an empirical estimate of the probability distribution $\hat{p}(\rho_\mathrm{true},\mathrm{POVM})$ \COMMENT \underline{No direct access to $\rho_\mathrm{true}$}
			\FOR{member of particle bank $\rho_i$}
			\STATE Calculate Born probabilities $p_i$ of measuring the $\mathrm{POVM}$ on $\rho_i$
			\STATE Calculate distances ($L_1,L_2$) between $p_i$ and $\hat{p}$.
			\STATE NN(distances) $\to$ single combined distance
			\STATE NN(distances) $\to \varepsilon_i$ \COMMENT perturbation size
			\STATE NN(distances) $\to$ $w_i$
			\ENDFOR
			\STATE $\sum_i \rho_i w_i \to \rho_\mathrm{guess}$ \COMMENT Best guess from this step
			\STATE NN(combined distances,weights) $\to$ guess score
			\STATE Draw new bank from bank at random based on weights.
			\FOR{step $s$ from 1 to $N_{\mathrm{steps}}$}
			\FOR{member of new particle bank $\rho_{i'}$}
			\STATE Purify $\rho_{i'} \to \vec{v}_{i'}$.
			\STATE Generate $N_\mathrm{resample}$ random vectors $\left \{ \vec{u} \right \}$.
			\FOR{random vector $\vec{u}$}
			\STATE Keep the orthogonal part $\vec{o}$ of $\vec{u}$ to $\vec{v}_{i'}$.
			\STATE Normalize $ \vec{o} / |\vec{o}| \to \vec{o}$
			\STATE Combine with purified bank particle as $\vec{v}_{i'}^\mathrm{perturbed} = (1-\varepsilon_{i'}) \vec{v}_{i'} + \varepsilon_{i'} \vec{o}$. \COMMENT $\varepsilon$ chosen by NN
			\STATE Depurify $\vec{v}_{i'}^\mathrm{perturbed} \to \rho_{i'}^\mathrm{perturbed}$
			\STATE Obtain outcome probabilities $p^\mathrm{perturbed}_{i'}$ of measuring the $\mathrm{POVM}$ on $\rho_{i'}^\mathrm{perturbed}$
			\STATE Calculate distances between $p^\mathrm{perturbed}_{i'}$ and $\hat{p}$.  
			\ENDFOR
			\STATE Keep the perturbed $\rho$ with the smallest distance.
			\ENDFOR
			\STATE New bank of particles $\to$ bank of particles
			\ENDFOR
			\STATE From the step $t$ we have ($\rho_\mathrm{guess}$, guess score, bank)
			\STATE Using guesses from steps $<t$, calculate $\text{guess weights} = \mathrm{softmax}(\text{guess scores})$.
			\STATE $\hat{\rho}_t = \sum_{t' \leq t} \text{guess weight}_{t'} \,  \rho_{\mathrm{guess},t'}$ \COMMENT Output guess
			\STATE $\mathrm{Loss}_t = \mathrm{Distance}(\hat{\rho}_t, \rho_\mathrm{true})$
			\IF{using measurement adaptivity} 
				\STATE Generate a set $A$ of random angles $(X,Y,Z)$ for each qubit.
				\FOR{choice of angles from the set of angles $A$}
					\STATE Create a POVM parametrized by the angles.
					\STATE Obtain outcome probabilities $p_{\mathrm{mean}}$ of measuring the POVM on the output guess $\hat{\rho}_t$
					\STATE For all members of the particle bank $i$ obtain outcome probabilities $p_i$ of measuring the POVM on the particle $\rho_i$.
					\STATE Calculate a mixed entropy heuristic $f = S(p_{\mathrm{mean}}) - \sum_i w_i S(p_i)$, where $S(p)$ is the entropy of distribution $p$. 
				\ENDFOR
				\STATE Choose the POVM corresponding to maximum $f$.
			\ENDIF
			\ENDFOR
			\STATE $\mathrm{Loss} = \sum_t \mathrm{Loss}_t$
		\end{algorithmic}
	\end{algorithm}
\end{figure} 

Figure \ref{fig: unit cell} shows the inner architecture of each RNN unit cell, which during training includes simulated quantum mechanical measurements. During testing and while deployed in a lab, the black box measurements are supplied by the experimenter. After every batch of simulated measurements on the true state, the empirical counts of measurement outcomes are collected. These are then compared to the probability distributions induced by the current POVM $\{\Pi_{\gamma}\}_{\gamma}$ on the candidate particles in the bank
$\left [ p_i = \mathrm{Tr}(\Pi_{\gamma}\rho_i)\}_{\gamma} \, , \forall i \in \mathcal{B} \right ] $
via the $L_1$ and $L_2$ distances on probability distributions. Based on the distances, a neural network generates a set of weights $\{w_i\}_{i \in \mathcal{B}}$ to associate with the corresponding particles, which are in turn used to produce a state reconstruction. In addition, the optimal perturbation size per particle $\varepsilon_i$ (for use in the resampling step later) as well as an overall confidence score in the reconstruction are also generated using independent neural networks, as described in detail in Algorithm~\ref{alg:NQST-train} and illustrated in Figure~\ref{fig: NN diagram}.

Before we go on to describe the technically more involved resampling and measurement adaptivity steps, we offer two comments. Firstly, NA-QST is guaranteed to output a valid density matrix as its running estimate of the state is always a convex linear combination of the `particles' -- which are themselves valid states -- with normalized weights. Second, unlike methods that rely on Bayesian weight updates (such as ABQT), our weights do not have the interpretation of being a posterior distribution on the particles. Indeed, the Bayesian, principled method of updating weights comes at a heavy computational price. In ABQT, the posterior distribution forms the basis of their adaptation function (see further details below), and an accurate representation of the posterior distribution requires many particles, and an accurate update of weights. However, as mentioned earlier, as one accumulates data in QST, the posterior distribution gets more and more sharply-peaked. As a result, in ABQT, most of the weights of an initial bank of particles decay to small values and become non-zero only for few particles.

The numerical fix for this is resampling, which consists in choosing a new set of particles out of the old set with probabilities proportional to their old associated weights, perturbing them slightly, and then assigning them uniform weights. Taking a leaf from this book, NA-QST also periodically resamples particles, but significantly faster and more effectively as the resampling procedure is learned directly from data. In our case the resampling is not a mere numerical fix (as we do not suffer from the weight decay problem) but the crucial inferential step, as the refinement of the particle bank allows the algorithm to achieve better precision in the reconstruction. 

The resampling proceeds as follows: from the previous step, every particle has an associated random perturbation magnitude $\varepsilon_i$, which represents the neural network's estimate of how far the $i$-th particle is from the true state. Each particle, generally a mixed state, is purified using a reference system of equal dimension and perturbed by $\varepsilon_i$ in $N_{\text{resample}}$ randomly-generated orthogonal directions according to the expression specified in Algorithm~\ref{alg:NQST-train} (working with purified states allows more effective perturbation). The perturbed particle candidate is the density matrix obtained after partial-tracing away the reference system (``de-purifying"). Of these $N_{\text{resample}}$ candidates, only one is retained: the one whose Born measurement probabilities are closest to the empirical distribution just measured. This procedure is repeated $N_\mathrm{steps}$ times. Overall, all particles are perturbed so that they move closer to the true state; these then make up the new bank. A numerical perspective on this is that we have effectively run an approximation to gradient descent on $\mathrm{distance}(p_{\text{particle}},p_{\text{empirical}})$, within the larger gradient descent training loop. This nesting is called \textit{meta-learning} \cite{pmlr-v70-finn17a} and it is an active area of machine learning research. 

Finally there is the measurement adaptivity step given by maximizing the function 

\begin{align} \label{eq:heuristic}
\arg \max_{\alpha \in \mathcal{A}} \left\{H{\bigl[p_y\bigl(\alpha,\hat\rho(\mathcal{D})\bigr)\bigr]} - \mathbb{E}_{q(\rho \lvert \mathcal{D})} {\left[H{\bigl[p_y\bigl(\alpha,\hat\rho(\mathcal{D})\bigr)\bigr]}\right]} \right\}.
\end{align}
$\mathcal{A}$ denotes a class of POVM considered for the adaptation procedure, and $\alpha$ labels a  member of that class. As before, $\mathcal{D}$ is the measurement data collected thus far. $p_y$ is the probability from Born's rule [Eq.~\eqref{eq:BornRule}], written here with the arguments $\alpha$, specifying the POVM used, and the state $\hat\rho$, symbolizing our current best estimate of the state given the data $\mathcal{D}$. $\mathbb{E}_q[\cdot]$ denotes the expectation value of the argument with respect to the distribution $q$. $q(\rho|\mathcal{D})$ abstractly denotes some probability distribution over the continuous state space based on the data $\mathcal{D}$; in practice, it is the discrete probability distribution as specified by the current bank $\mathcal{B}$ of particles $\rho_i$, with their weights $w_i$. We use $\hat\rho\equiv \sum_{i\in\mathcal{B}}w_i\rho_i$, which depends on the data through the adjustment of the bank as data is gathered. Note that this is the same function as used in the original ABQT paper \cite{HuszarAdaptiveTomography}, except that they have $q(\rho|\mathcal{D})\equiv p(\rho|\mathcal{D})$ the posterior distribution. In NA-QST, the distribution $q$ is determined by the bank chosen by the neural network.

\begin{figure*}
 \includegraphics[width=0.3\linewidth]{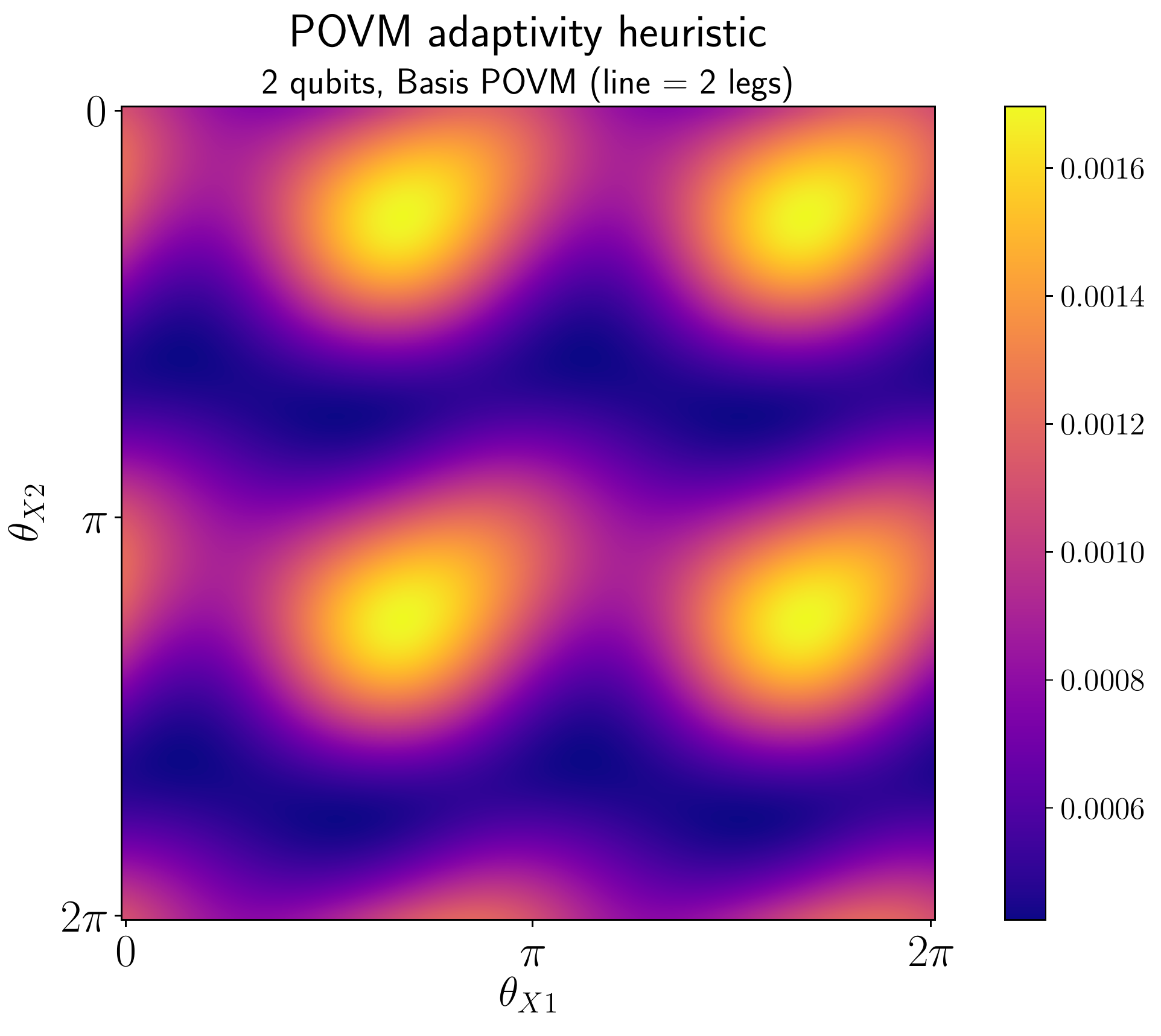}%
 \includegraphics[width=0.3\linewidth]{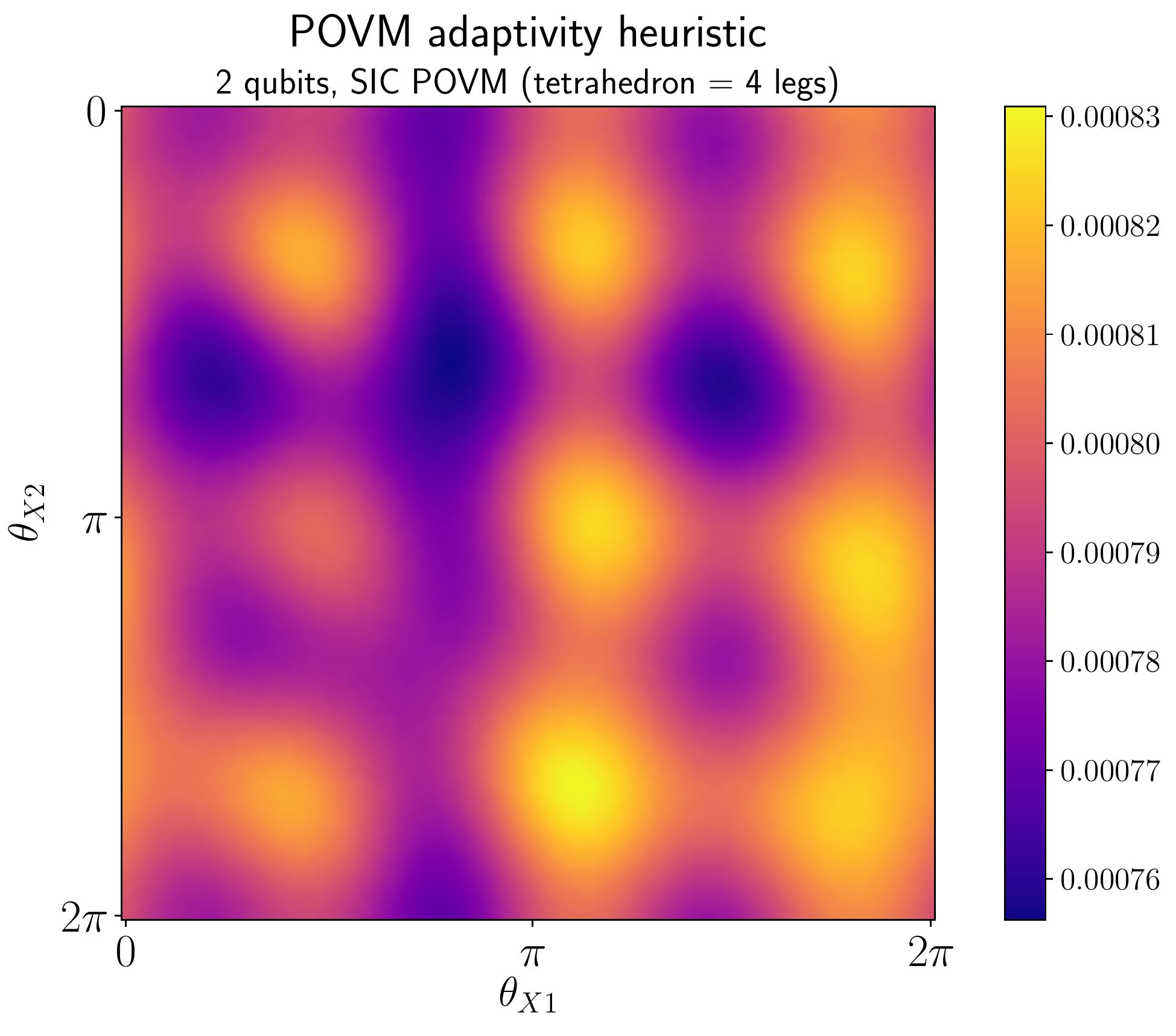}%
 \includegraphics[width=0.3\linewidth]{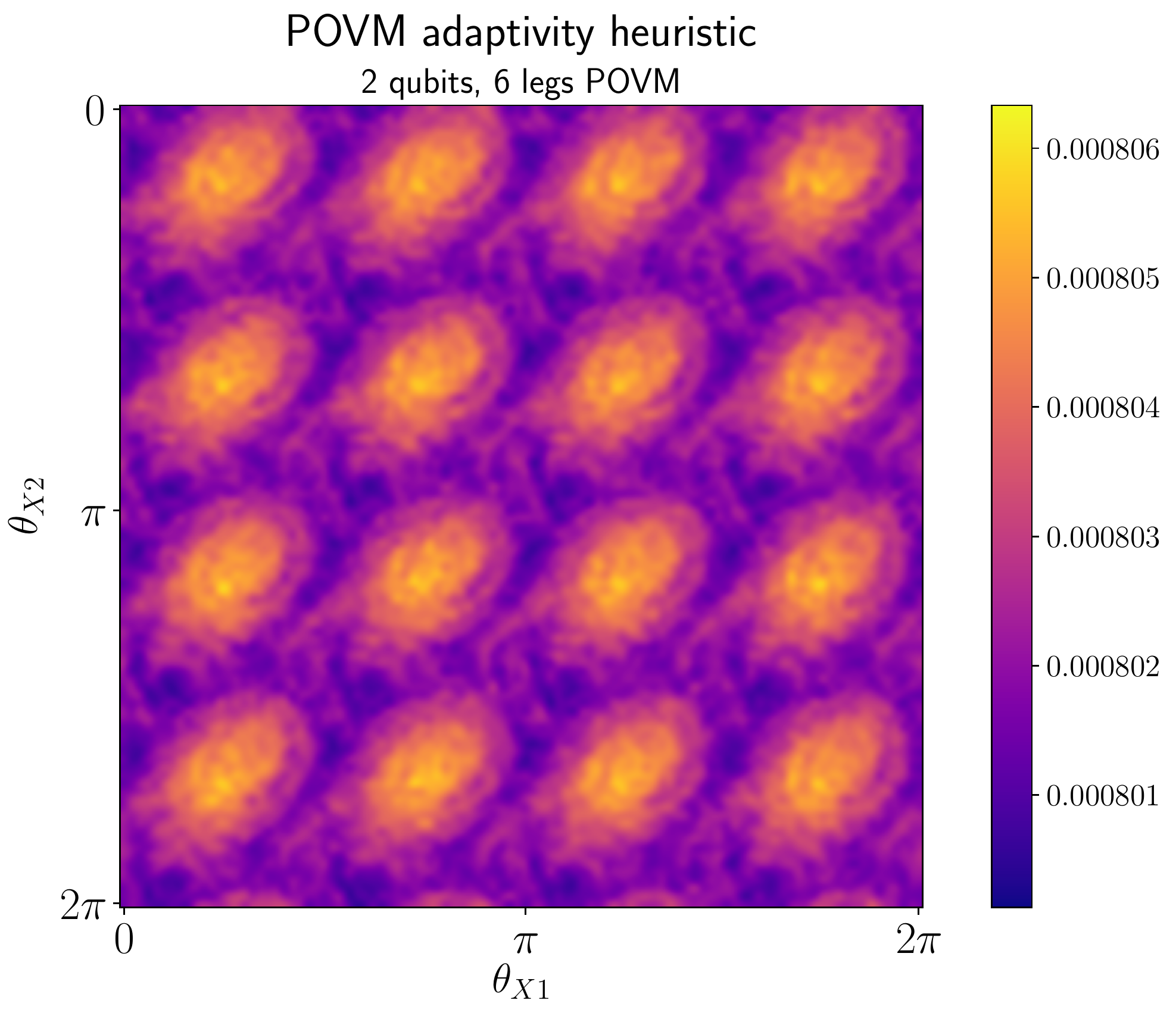}%
 \caption{\label{fig:entropy_heuristic} The entropy-based heuristic used to choose the optimal POVM angles. The figures show examples of two-dimensional sections of the entropy-based heuristic that we use to choose the POVM angles parametrizing its orientation in the respective qubit subspaces. The figures show an example for a basis POVM (two legs lying in opposite directions for each qubit), SIC POVM (4 legs in the shape of a tetrahedron for each qubit) and a POVM with 6 legs in $\pm X$,$\pm Y$, and $\pm Z$ directions for each qubit. The range of values attained in the section for 4 and 6-legged POVM is negligible compared to the basis POVM. This is due to the fact that they, unlike the basis POVM, are informationally complete, i.e. provide similar amount of information regardless of their orientation. The x and y axis are the $X$-axis rotation angles for the two qubit subspaces.}
 \label{fig:heuristic}
 \end{figure*}

The motivation for choosing this function is explained in the Appendix. In essence, we are looking for a POVM specified by a set of angles such that the algorithm's ability to distinguish between particles increases, as well as its ability to measure distances to the true density matrix. We experimented with a large number of approaches, including analytical gradient descent, however, the problem proved to be best solved by a simple approach: choose a random set of POVM angles, calculate the heuristic in Eq.~\eqref{eq:heuristic} from the previous step, repeat several times (we chose 40 as a good trade off between accuracy and speed), and keep the POVM that maximizes the heuristic. We experimented with different choices of the heuristic as well as learning it from data, however, as demonstrated by empirical experiments summarized in Figure~\figref{fig:heuristic}, the function Eq.~\eqref{eq:heuristic} was simple enough for our random method to work well. 

We round off this section with some remarks. The basic mechanism employed by NA-QST differs quite significantly from that of ABQT. In proposing new weights and perturbations, the neural network does so according to its own, learned, similarity metric between the particles and the true state, based on the distances between the measurement probabilities of each particle and the empirical measurement statistics. In ABQT, by contrast, the size of the perturbation is chosen from a fixed, approximately Gaussian distribution (see \cite{Struchalin2016ExperimentalStates}). This is because in ABQT, the perturbation merely provides variation to the particles in the bank, whereas our perturbations are expressly chosen to bring the particles as close as possible to the current estimate of the state. Secondly, our algorithm does not merely concatenate the disjoint weight updates/resampling and measurement adaptivity steps; by putting them both in the same neural architecture, the measurement adaptivity heuristic influences training such that the weights output by the former two upstream tasks are well-suited for subsequent input into the latter heuristic.

\section{\label{methods} Methodology}

As our main benchmark for the performance of our adaptive neural network based algorithm, we use ABQT, the adaptive algorithm first proposed in \cite{HuszarAdaptiveTomography} and computationally refined in \cite{Struchalin2016ExperimentalStates}. As mentioned earlier, the core idea of the ABQT algorithm is to keep a bank of samples from the state space and construct a posterior on this bank. The posterior is updated in a Bayesian fashion with every new batch of measurements. Subsequently, the measurement configuration for the next batch is chosen by optimizing a heuristic given by  Eq.~\eqref{eq:heuristic}. The final state estimate is the weighted average of this bank. 

We reimplemented the ABQT algorithm in the Python~3 programming language to be able to compare the runtime with our neural network algorithm, written in TensorFlow and Python. We verified our implementation against results in the original paper.

Following \cite{Struchalin2016ExperimentalStates}, we consider 2-qubit examples. The POVMs we consider here are \emph{product} POVMs, i.e., $\Pi\equiv \Pi^{(1)}\otimes\Pi^{(2)}\equiv \{\Pi_{y_1}^{(1)}\otimes \Pi_{y_2}^{(2)}\}$, where $\Pi^{(i)}$ are 1-qubit POVMs. Product POVMs are typically the easiest to implement in experiments. In particular, we assume that $\Pi^{(1)}$ and $\Pi^{(2)}$ are the same type of 1-qubit POVM, although, we allow for the orientation of the POVM outcome operators for each qubit to be chosen independently (see details below). We follow \cite{Struchalin2016ExperimentalStates} in using the squared Bures distance between the true state $\rho$ and the estimate~$\hat{\rho}$ 
\begin{equation*}
d_B^2(\rho,\hat{\rho}) = 2-2 \sqrt{F(\rho,\hat{\rho})}
\end{equation*}
as our figure of merit for accuracy of reconstruction. Here, $F(\rho,\sigma) \triangleq \mathrm{Tr} \left[ \left(\sqrt{\rho^{\frac{1}{2}} \sigma \rho^{\frac{1}{2}}} \right) \right] ^2 $ is the squared-fidelity  between two states $\rho$ and $\sigma$. The Bures distance between two states has an operational interpretation as the maximum possible Kullback-Leibler (KL) divergence between output statistics of the same set of quantum measurements on the two states. Intuitively, a large Bures distance between two states indicates that they are \textit{easy} to distinguish using quantum measurements, and vice versa \cite{Fuchs1995MathematicalTheory}. Although the Bures distance was the evaluation metric for reconstruction accuracy at test time, for training we used the Hilbert-Schmidt, or $L_2$, distance as the loss function, for computational stability reasons. The choice of Bures or $L_2$ distance in the loss function had no observable effect on the trained solution. For small distances, both metrics are very similar. 

\begin{figure}[ht]
\includegraphics[width=1.0\linewidth]{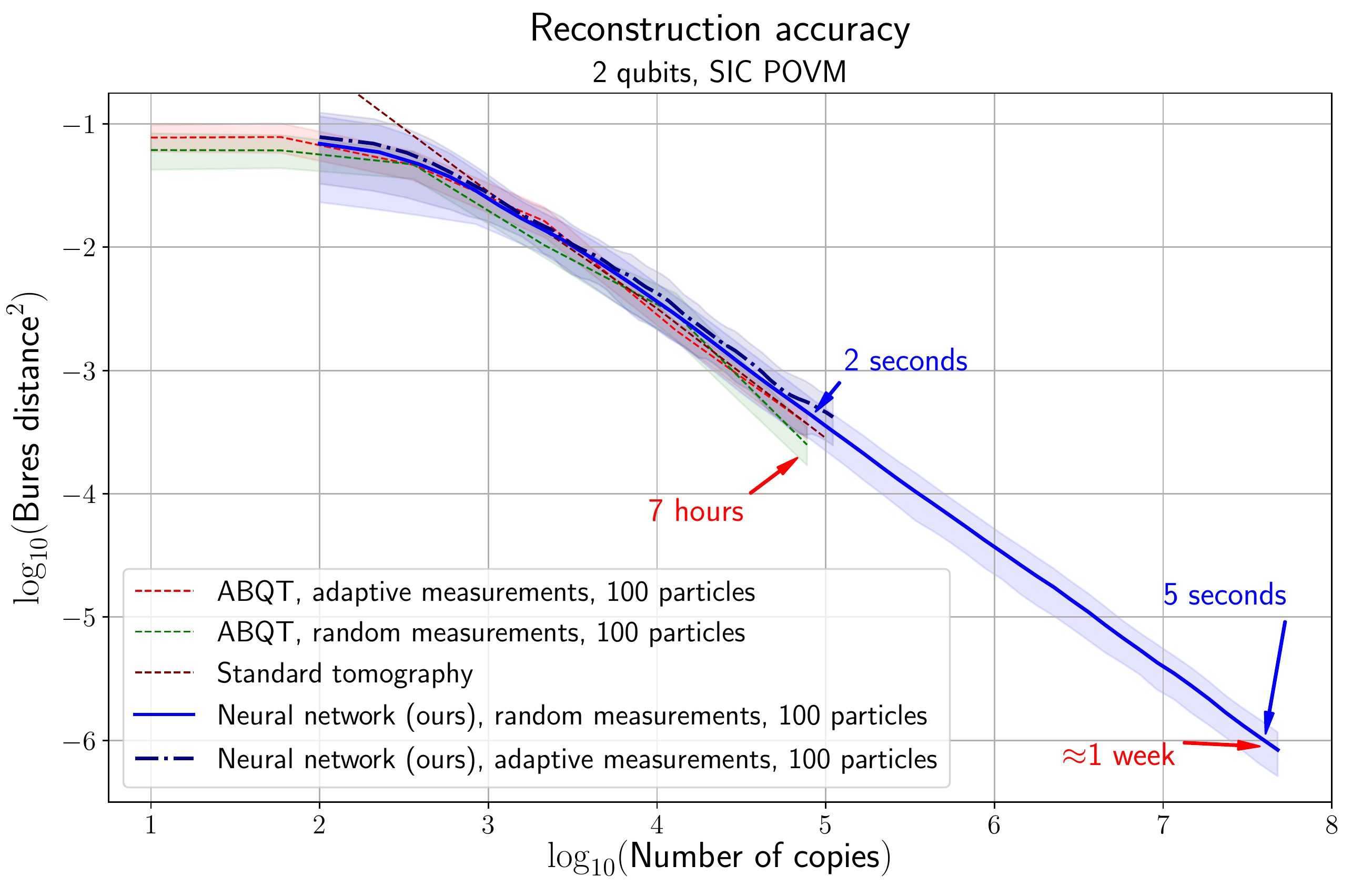}
\caption{Performance comparison for algorithms using the 1-qubit SIC-POVMs (4 legs arranged in a tetrahedron for each qubit). Our neural network algorithm NA-QST is compared against ABQT and Standard QST. Given a number of copies measured, our neural algorithm performs equally to ABQT with both random and adaptive SIC-POVMs (see definition of `random' POVMs in main text), as well as Standard QST. In addition, the required analysis runtime on a laptop favors NA-QST by a factor of $10^6$ for $10^7$ measurements (several seconds for our algorithm compared to a week (the bottom-right label) for ABQT per reconstruction). Comparing the performance of all our algorithms, we conclude that measurement adaptivity does not make a significant difference in performance using SIC-POVMs.}
\label{fig: performance comparisons}
\end{figure}
\begin{figure}
 \includegraphics[width=1.0\linewidth]{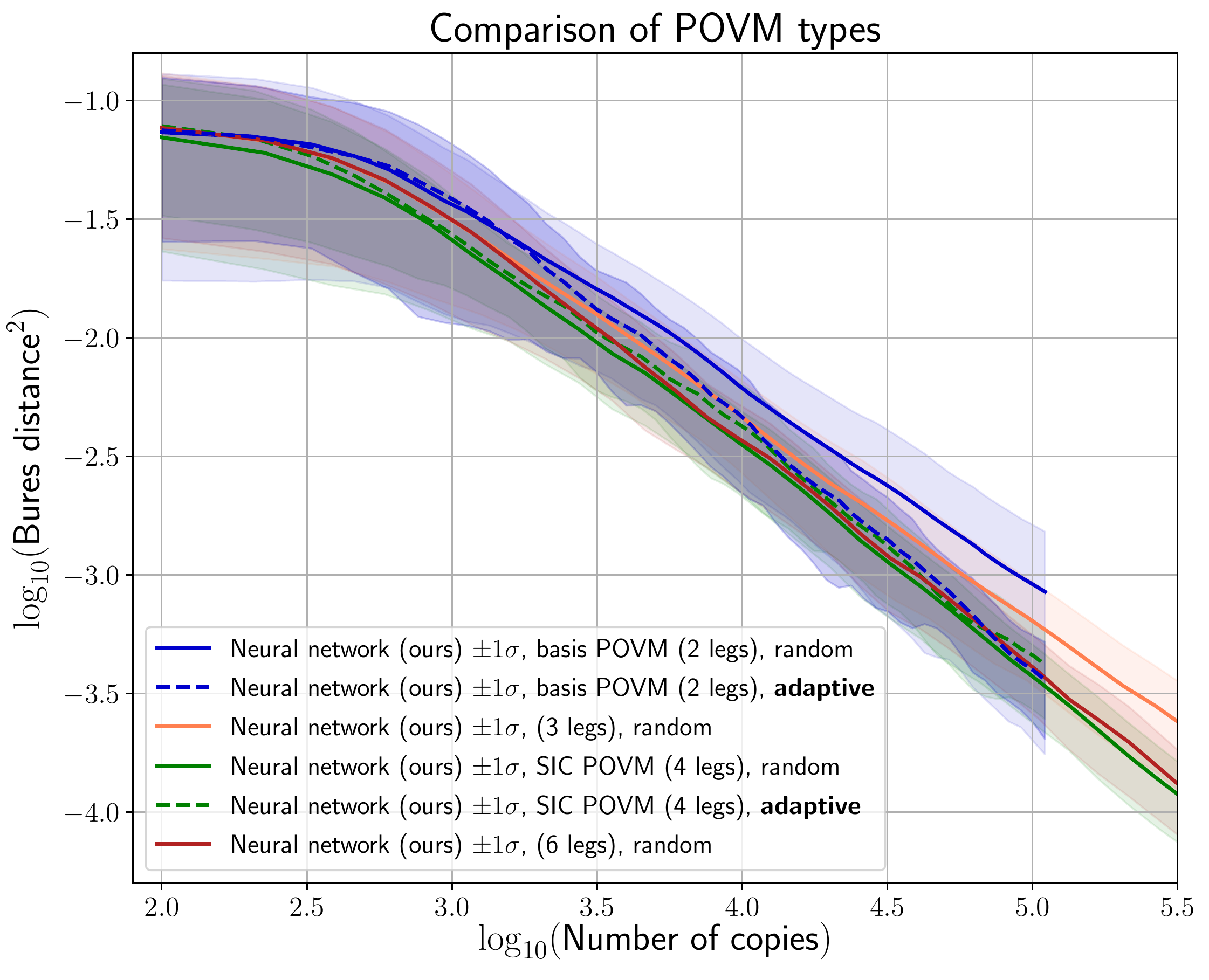}%
 \caption{\label{fig: 2346legs} Comparison of our trained NA-QST algorithms using different POVM types and measurement adaptivity. The accuracy of reconstruction is shown as a function of the number of available measurement outcomes. The shading is a $1 \sigma$ confidence interval. SIC (4 legs), 6 legs POVMs offer virtually identical performance and the effect of measurement adaptivity on them is negligible/non-existent. This is most likely due to the fact that they cover the whole space evenly and the POVM orientation therefore does not affect state distinguishability significantly. The basis POVM with random measurements performs significantly worse than the rest on average, and also has a wider uncertainty range. This is probably due to the effect of randomness on the correct POVM orientation. By using measurement adaptivity, our algorithm with basis POVMs performs as well as informationally complete POVMs. It seems that the effect of measurement adaptivity is to make the basis POVM informationally complete over several subsequent measurement steps.}
 \end{figure}

The parameters of variation that we explored are:

\begin{enumerate}

    \item Reconstruction algorithm. 
    
    Our main benchmark is ABQT, but we also implemented what we will refer to as Standard Quantum State Tomography (see Chapter 8 of \cite{NielsenQuantumInformation}.) Since this method is not guaranteed to give a valid density matrix, we had to implement an additional projection step, following the algorithm of \cite{Smolin2012} to project the output of this algorithm onto the set of valid density matrices.
    
    \item Type of 1-qubit POVM.
    
    We considered several different types of 1-qubit POVMs:
    \begin{enumerate}
    \item Basis POVMs (i.e., projective measurements).  Projective measurements, i.e., measurements of a single basis of states (two states for the qubit), are usually the easiest to implement in any experiment. Unfortunately, a single projective measurement is not informationally complete, i.e., one cannot reconstruct the full quantum state just by measuring a single basis all the time. This hints at a possible advantage of an adaptive procedure, where the orientation of the basis is chosen adaptively as one learns more about the state. This was done in \cite{Struchalin2016ExperimentalStates}, and we explore the example using NA-QST. Here, we examined the performance using 5, 30 and 100 particles in the bank for both the adaptive and random (see below) ABQT. Our neural algorithm used only 100 particles and we experimented with both random and adaptive measurements. The results are shown in Figure~\ref{fig:2legs}. 
    \item The one-qubit symmetric informationally complete (SIC) POVM, i.e., the tetrahedon POVM \cite{MinimalTomo}. The tetrahedron POVM is thus named because its four outcomes can be represented in the qubit Bloch sphere as four legs extending from the centre of the sphere to the vertices of a regular tetrahedron. Here, we allow the tetrahedron orientation to be chosen by the algorithm. The results of comparing all algorithms restricted to using this POVM are shown in Figure~\ref{fig: performance comparisons}. 
    \item One-qubit POVMs with three or six legs per subspace -- to see the performance of NA-QST for different POVM types.  The orientation of the legs relative to each other is fixed; only the overall orientation is allowed to vary. The comparison of reconstruction accuracy for NA-QST using POVMs with different number of measurement outcomes is presented in Figure~\ref{fig: 2346legs}.
    \end{enumerate}
    
    \item Adaptive or random measurements. 

    \textit{Random} refers to the possibility of switching off measurement adaptivity -- that is, to choose all measurement POVMs randomly before the start of the experiment, as opposed to choosing them adaptively with sequential measurements. \textit{Adaptive} measurements are chosen by our algorithm according to Eq.~\eqref{eq:heuristic} as the suitable choice for the next set of measurements, implicitly maximizing the information gained from it. The results of turning adaptivity on and off for the adaptive algorithms are shown in Figure \ref{fig:2legs}.
    
    \item Size of particle bank in ABQT.
    
    We varied the size of the bank of particles to use 5, 30 and 100 particles and examined the impact on performance. Since in ABQT, the bank consists of samples of the posterior distribution used to compute the adaptivity heuristic [Eq.~\eqref{eq:heuristic}], finer sampling should improve performance but comes at a computational cost. Our computational power limited us to using no more than 100 particles for ABQT as well as NA-QST.
\end{enumerate}
 
\section{Results}

\subsection{\label{results1} When is adaptivity helpful? }

Using our algorithm's rapid training capabilities, we can shed some light on when adaptivity is useful, for the variables considered here. In all our comparisons mentioned below, the metric we use for `efficacy' is the scaling of accuracy of reconstruction (measured by Bures distance from the estimate $\hat{\rho}$ to the true $\rho$) with the number of measurements.

\begin{figure}[htbp!]
 \includegraphics[width=1.0\linewidth]{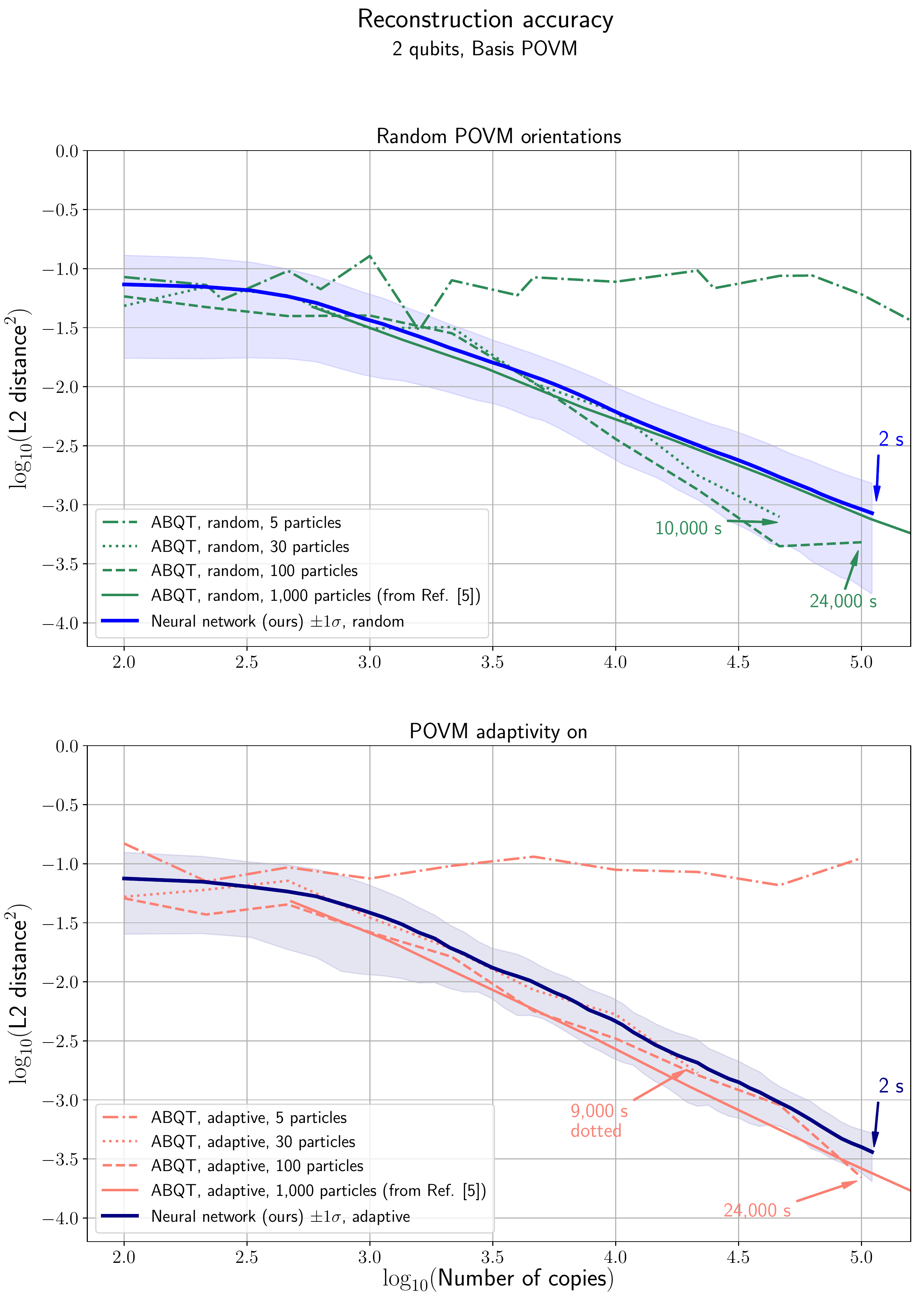}%
 \caption{Performance comparison for algorithms using Basis POVMs (2 legs arranged on a line for each qubit). Our neural network algorithm is compared against ABQT ran with 5, 30 and 100 particles. Given a number of copies measured, our neural algorithm performs comparably (within $1 \sigma$) to ABQT with both random and adaptive POVMs. The approximate runtimes to a given accuracy are marked on the plots, strongly favoring our algorithm by a factor of several thousand. Our neural architecture learned to approximate time-consuming Bayesian calculations using equally powerful and fast-to-compute heuristics. Measurement adaptivity significantly improves the reconstruction accuracy for both NA-QST as well as AQST.}
 \label{fig:2legs}
 \end{figure}

Our first comparison in Figure \ref{fig: performance comparisons} fixes the POVM to be an SIC (4 legs per qubit) POVM. We then examine the performance of various reconstruction algorithms. Five algorithms were compared: 1) Standard Quantum State Tomography; ABQT (as in Refs.~\cite{HuszarAdaptiveTomography} and \cite{Struchalin2016ExperimentalStates}) with 2) adaptively- and 3) randomly-chosen measurement configurations; NA-QST with 4) adaptively and 5) randomly-chosen measurement configurations. Effectively, the performances of all algorithms are identical, indicating that when SIC-POVMs are used, adaptivity yields no advantage over standard tomography. Another interesting contrast shown in Figure \ref{fig: performance comparisons} is that between random and adaptive measurements in the adaptive algorithms. Essentially, adaptivity yields no benefit over random measurement configurations when an SIC-POVM is used. The situation is different in Figure \ref{fig:2legs}, where projective measurements (2 legs) are used. Projective measurements are much less informative than SIC-POVMs, and so, the ability to adapt the POVM significantly improves performance.

Our second comparison in Figure \ref{fig: 2346legs} further explores the effect of POVM type on performance of the NA-QST algorithm. We ran NA-QST on POVMs with different numbers of legs per qubit -- 2 (Basis POVMs, subtending a line through the centre in the Bloch sphere), 3 (the trine measurement, a flat equilateral triangle), 4 (SIC-POVMs, a regular tetrahedron) and 6 (the standard six-state POVM measuring the three Pauli operators, a diamond). The performance is effectively identical for the last two, and is significantly worse for basis POVMs without measurement adaptivity. It appears to make no difference what type of POVM we use beyond using an informationally complete one ($\ge 4$ legs). The effect of measurement adaptivity on basis POVM (2 legs) seems to effectively turn it into an informationally complete POVM over several measurement steps. 

Finally, we offer some numerical/empirical evidence to back up our claim that adaptivity is unhelpful for SIC POVMs and beyond: the range of variation of the heuristic function [Eq.~\eqref{eq:heuristic}], optimized in the adaptive step, is drastically smaller ($2-3\%$ of maximum for SIC-POVMs as compared to about $75\%$ for basis POVMs). As we discuss in the Appendix, this heuristic function can, in fact, be identified with a quantity known as the {\em accessible information} of the ensemble of states in the bank, which represents the maximal mutual information between the random variable storing the next POVM measurement result, and the label of the bank particles -- a random variable distributed according to the bank weights/posterior distribution. The fact that it does not vary much when the optimization is over the class of SIC-POVMs is an indication that the amount of information yielded by SIC-POVMs is almost invariant to their actual orientation.

\begin{figure}[htbp!]
        \includegraphics[width = 1.0\linewidth]{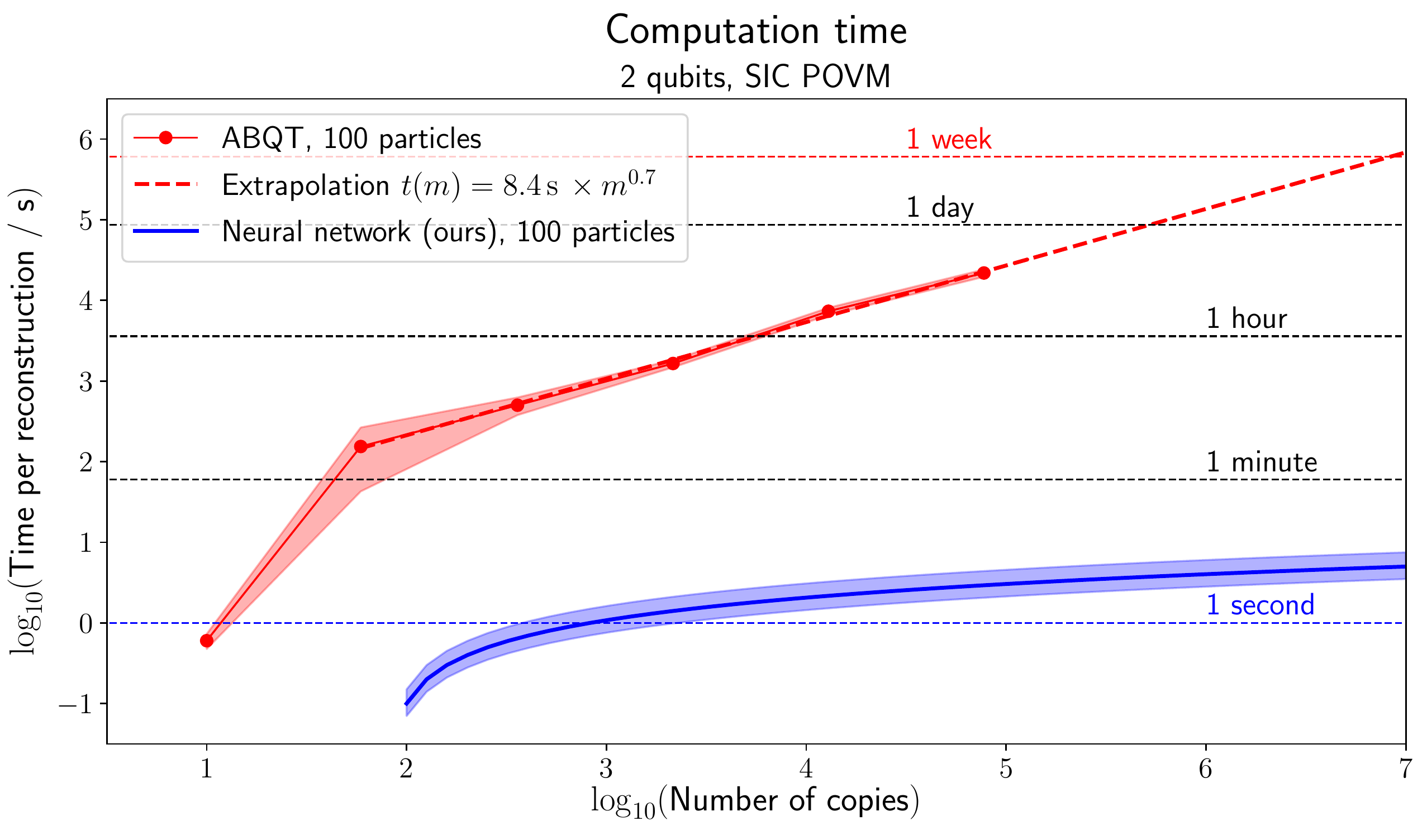}
        \caption{\label{fig: runtime_comparison} Scaling of reconstruction runtime with number of copies measured. Our neural algorithm runtime empirically scales \textit{logarithmically} with the number of copies measured, while ABQT empirically scales \textit{polynomially} (approximately as $m^{0.7}$). We ran comparisons on the same hardware. However, for numbers of measurements $>10^5$ we had to extrapolate the ABQT runtimes, since it was taking tens of hours per reconstruction. For $10^7$ measurements, our algorithm performs a reconstruction in $\approx$5~seconds, while the ABQT would take approximately a week (extrapolated).} 
\end{figure}
\subsection{\label{results2} Entropy heuristics for POVMs}

To choose the POVM that will maximize information gained in the next measurement, we investigated a number of heuristics that could guide the selection of the measurement angles. We also attempted learning such a function from data. However, we found that what works best is the accessible information heuristic in Eq.~\eqref{eq:heuristic}. We optimize the POVM angles by choosing them at random and selecting the one attaining the highest value of the heuristic. This simple approach performed better than more advanced tools such as meta-learning which we also experimented with.

\subsection{\label{results3} Adaptivity with Basis POVMs: accuracy and runtime}

Based on the conclusions of the previous part, we specialize to basis POVMs to showcase our adaptive algorithm in this subsection. We emphasize again that basis POVMs are also the easiest to implement in the laboratory.

Here we explicitly compare the performance of our NA-QST in terms of reconstruction accuracy to ABQT. Figure \ref{fig:2legs} shows this comparison, using both random and adaptive measurements. Generally, ABQT improves as more particles are used -- since better estimates can be made when the posterior is more finely sampled. If the number of particles is fixed at 100 for both NA-QST and ABQT, the performance of NA-QST is comparable to ABQT for both types of measurements. In our experiments, performance of NA-QST improves with more particles as well. For computational reasons (limited memory on a laptop), we could not efficiently explore more than 100 particles for NA-QST. ABQT's performance seems to empirically plateau for 30 particles and above. 

Our second comparison examines the scaling of computational runtime with the number of measurements to analyze. Figure \ref{fig: runtime_comparison} shows that ABQT takes prohibitively long runtimes that are polynomial (to the power of $\approx0.7$ according to our experiments) in the number of measurements. This is because the resampling step requires computation of a full posterior over all previous measurements. By contrast, the runtime of NA-QST seems to be logarithmic in the number of measurements. For $10^7$ measurements to analyze, our NA-QST converges to a solution in approximately 5 seconds on a laptop, whereas the ABQT would take approximately a week to run.


\section{Conclusions}

Overall, the neural network based algorithm for quantum state tomography (NA-QST) we developed and ran in Figures~\ref{fig: performance comparisons}, \ref{fig: 2346legs} and \ref{fig:2legs} performs comparably to ABQT with both adaptive and random measurements, but with a significant reduction in computational time, as evidenced in Figure \ref{fig: runtime_comparison}. This is due to our fast implementation of the resampling step which avoids the computational bottleneck of computing a full posterior and instead learns an effective heuristic directly from data. In the same amount of time, ABQT is able to keep track of much fewer samples of the posterior, which worsens its performance significantly. Furthermore, by training the resampling and adaptive components of our algorithm jointly using the same loss function, the resampled weights chosen by the neural network are near-optimal choices for the ensemble on which the adaptive heuristic is calculated. This allows for rapid adaptation to particular experiment's needs, as we can retrain within a few hours on a laptop.

Our work is part of an early wave of investigations that use neural networks to improve algorithms in experimental physics that rely on some aspect of signal processing and control theory. In quantum tomography specifically, efficacy-enhancing neural networks have also started to gain recognition. The work of  \cite{Torlai2018Neural-networkTomography} to reconstruct a many-body quantum wavefunction using Restricted Boltzmann Machines is one such example, although it is different both in terms of the machine learning architecture used and the desired outcome. One advantage of neural networks is their versatility; in this paper we have tackled estimation of density matrices of fully-general states, but if one has prior knowledge about the state to be estimated (for instance if it is known that the state is a low-rank state \cite{Gross2010QuantumSensing,Flammia2012QuantumEstimators} or a Choi matrix for a particular class of quantum channels), it is a simple matter to retrain our algorithm only on that specific class of states, allowing it to specialize to it. A further improvement could be achieved by parameterizing the neural network's output appropriately. Alternatively, one might not necessarily be interested in a full description of the state's density matrix, but only functions of said matrix, such as entanglement entropy. For these purposes, our algorithm would only require minor modifications and would be quick to retrain.

A hurdle to full quantum tomography of high-dimensional states is the fact that computational complexity grows exponentially in the number of qubits to be estimated. This is unavoidable in any tomography algorithm, since the dimension of the required output itself grows exponentially. However, neural networks and machine learning in general could be the key to achieving a favorable pre-factor for this exponential complexity, which would push the current tractable dimension boundary even higher. Our work shows a very significant runtime speedup that we intend to use to investigate systems $> 2$ qubits. 

In this paper, we demonstrated the power of learning heuristic solutions to computationally complex problems directly from data. By doing so, we were able to develop a quantum state tomography algorithm that achieves orders of magnitude faster runtime while retaining state-of-the-art reconstruction accuracy. As a by-product, we implemented a fully differentiable (in the machine learning sense) version of many quantum mechanical tools, which can be reused in other machine learning solutions to quantum mechanical problems. We further show that measurement adaptivity is only of significant help for basis POVMs, as the adaptivity improves the reconstruction accuracy to the level of other informationally-complete POVMs. We are planning to publish the core part of our code.

\subsection*{Acknowledgments} 
S.F. was supported by a Research Assistantship at the Kavli Institute for Particle Astrophysics and Cosmology at Stanford University. Y.Q. was supported by a Stanford (Gabilan) Graduate Fellowship and a National University of Singapore Overseas Graduate Scholarship. This work is funded by the Singapore Ministry of Education (partly through the Academic Research Fund Tier 2 MOE2016-T2-1-130) and the National Research Foundation of Singapore. H.K.N is also supported by Yale-NUS College through a start-up grant.

\appendix

\section{Interpretation of adaptivity heuristic}
We offer an interpretation of the heuristic used for measurement adaptivity (Equation~\ref{eq:heuristic}). 
A well-known quantity in quantum Shannon theory is the accessible information (see for example the textbook \cite{Wilde2011FromTheory}), which has the following operational interpretation: suppose Alice prepares a state by sampling from the ensemble $\varepsilon = \{p_X(x), \rho_x\}_{x\in\mathcal{X}}$, and gives the state to Bob. Bob does not know the value of the label $X$ (which is now a classical random variable with the distribution $p_X$), but is allowed to measure this state using a POVM $\{\Pi_y\}_y$.  If his outcome is stored in a random variable $Y$, the accessible information is the maximum (over all measurements) amount of information that $Y$ can yield about $X$:
\begin{align}
    I_\text{acc}(\varepsilon) = \max_{\{E_y\}} I(X:Y) \, ,
\end{align}
where $I(X:Y) \triangleq H(X) + H(Y) - H(X,Y)$ is the classical mutual information. 

Coming back to adaptive tomography, we now show that the adaptivity heuristic (Equation~\ref{eq:heuristic}) evaluated at its maximizing POVM is in fact $I_\text{acc}(\varepsilon)$ for $\varepsilon$ the ensemble consisting of the bank of particles and corresponding weights as relative probabilities.

Defining the following probabilities evaluated at the maximizing POVM, $\Pi^*_y$:
\begin{align}
    p_{Y\lvert X}(y\lvert x) &\triangleq p_X (x) \Tr (\Pi^*_y \rho_x)\\
    p_{Y\lvert \varepsilon}(y) &\triangleq \Tr (\Pi^*_y \sum_{x\in \mathcal{X}} p_X(x) \rho_x) = \sum_{x\in \mathcal{X}} p_X(x) p_{Y\lvert X}(y\lvert x) \label{eq:cond}
\end{align}
The mutual information evaluated at the maximizing POVM is
\begin{align}
    I_{\text{acc}}(\varepsilon) 
    &\triangleq \sum_{x,y} p_{X,Y}(x,y) \log \frac{p_{X,Y}(x,y)}{p_{X}(x)p_Y(y)} \nonumber \\
    & = \sum_{x,y} p_X(x) p_{Y\lvert X}(y\lvert x) \log \frac{p_{Y\lvert X}(y \lvert x)}{p_Y(y)} \label{eq:MI} 
\end{align}

Identifying the label of the particle in the bank with $X$, the posterior probabilities/weights with $p_X(x)$ and the optimal POVM's measurement outcome $\gamma$ with $Y$, the maximal value of Equation~\ref{eq:heuristic} is
\begin{align*}
&\mathbf{H}_Y\left[p_{Y\lvert \varepsilon}(\cdot)\right] - \sum_{x\in \mathcal{X}} p_X(x) \mathbf{H}_Y \left[p_{Y\lvert X}(\cdot \lvert x)\right]\\
&= -\sum_y p_{Y\lvert \varepsilon}(y) \log p_{Y\lvert \varepsilon}(y) \\
& + \sum_{x} p_X(x) \left(p_{Y\lvert X}(y\lvert x) \sum_{y} \log p_{Y\lvert X}(y\lvert x) \right)
\end{align*}
Using Equation~\ref{eq:cond} and the fact that $p_{Y\lvert \varepsilon} (y) = p_Y(y)$, we recover Equation~\ref{eq:MI}.

This interpretation, where we associate Alice's \textit{ensemble} with the particle bank and the experimenter's POVM choice with Bob's next measurement, yields some insight into the interplay of the adaptivity and resampling steps in ABQT: in between Bob's adaptive measurement choices, Alice's ensemble gets refined due to posterior updates and resampling. This also answers the question of why the resampling step of NA-QST -- which simply moves particles closer to the true state -- works: the refinement of the ensemble makes the adaptivity heuristic $I_\text{acc}(\varepsilon)$ calculated on the ensemble more salient to the state tomography task, yielding more information about the true state.

More concretely, in the setting we described, Alice had randomly chosen one of the states in the ensemble to be the true state, and $X$ is the label of the state. However, in ABQT, there is no obligation for the particle bank to contain the  true $\rho$ we wish to estimate (although it may contain close-by states). In fact, generically it does not, unless the ensemble is continuous over all of state space (i.e. infinitely many particles in the bank) and therefore contains $\rho$ by default. Therefore, with the discrete sampling of state space needed to form the ensemble $\varepsilon$ in ABQT, the adaptivity heuristic $I_{\text{acc}}(\varepsilon)$ is at best an approximation to $I_{\text{acc}}(\hat{\varepsilon})$, where $\hat{\varepsilon}$ is an ensemble that contains the true state. Furthermore, the more coarse-grained the sampling (i.e. the fewer particles in the bank), the poorer the approximation. The closer the particles cluster around the true state (as is the result of resampling) and more sharply-peaked the weights about the particles closest to the true state (as is the result of weight updates), the better the approximation.

\bibliography{references} 

\bibliographystyle{apsrev4-1}

\end{document}